\documentclass[fleqn,11pt,twoside]{article}

\usepackage{amsthm,amsthm,amssymb, color, xcolor,epsfig, graphics, subfigure}

\usepackage{amsmath, graphicx, latexsym, lscape }

\usepackage{cite}

\makeatletter
\newcommand{\copyrightnote}[2]{{\renewcommand{\thefootnote}{}
 \footnotetext{\small\it
\begin{flushleft}
 \copyright \ #1   #2
\end{flushleft}}}}

\newcommand{\Name}[1]{\begin{flushleft}
                       \LARGE \bf #1
                       \end{flushleft}\vspace{-3mm}}

\newcommand{\Author}[1]{\begin{flushleft}
                       \it #1 \end{flushleft}}

\newcommand{\Address}[1]{\begin{flushleft}
                       \it #1 \end{flushleft}}

\newcommand{\Date}[1]{\begin{flushleft}
                      \small  \it #1 \end{flushleft}}

\newcommand{\evenhead}{Author \ name}
\newcommand{\oddhead}{Article \ name}

\renewcommand{\@evenhead}{
\hspace*{-3pt}\raisebox{-15pt}[\headheight][0pt]{\vbox{\hbox to \textwidth
{\thepage \hfil \evenhead}\vskip4pt \hrule}}}
\renewcommand{\@oddhead}{
\hspace*{-3pt}\raisebox{-15pt}[\headheight][0pt]{\vbox{\hbox to \textwidth
{\oddhead \hfil \thepage}\vskip4pt\hrule}}}
\renewcommand{\@evenfoot}{}
\renewcommand{\@oddfoot}{}

\long\def\@makecaption#1#2{%
  \vskip\abovecaptionskip
  \sbox\@tempboxa{\small \textbf{#1.}\ \ #2}%
  \ifdim \wd\@tempboxa >\hsize
    {\small \textbf{#1.}\ \ #2}\par
  \else
    \global \@minipagefalse
    \hb@xt@\hsize{\hfil\box\@tempboxa\hfil}%
  \fi
  \vskip\belowcaptionskip}

\newcommand{\JNMPnumberwithin}[3][\arabic]{%
  \@ifundefined{c@#2}{\@nocounterr{#2}}{%
    \@ifundefined{c@#3}{\@nocnterr{#3}}{%
      \@addtoreset{#2}{#3}%
      \@xp\xdef\csname the#2\endcsname{%
        \@xp\@nx\csname the#3\endcsname .\@nx#1{#2}}}}%
}

\newcommand{\resetfootnoterule} {
  \renewcommand\footnoterule{%
  \kern-3\p@
  \hrule\@width.4\columnwidth
  \kern2.6\p@}
}

\renewcommand{\footnoterule}{}

\makeatother

\theoremstyle{definition}

\def\a{\alpha}
\def\b{\beta}
\def\ga{\gamma}
\def\de{\delta}   

\def\vphi{\varphi}

\def\vr{\varrho}

\def\s{\sigma}
\def\z{\zeta}

\def\vphi{\varphi}

\def\De{\Delta}
\def\La{\Lambda}

\def\pa{\partial}

\def\xb{{\bf x}}

\def\wb{{\bf w}}

\def\o+{\oplus}

\def\<{\langle}
\def\>{\rangle}

\def\({\left(}
\def\){\right)}
\def\[{\left[}
\def\]{\right]}
\def\=#1{\bar #1}
\def\~#1{\widetilde #1}
\def\wt#1{\widetilde #1}
\def\.#1{\dot #1}
\def\^#1{\widehat #1}

\def\"#1{\ddot #1}

\def\eeq{\end{equation}}
\def\beq{\begin{equation}}

\def\beql#1{\begin{equation} \label{#1}}

\def\eqref#1{(\ref{#1})}

\def\EOP{ \hfill $\triangle$ \medskip}

\def\symmref{AVL,CGbook,KrV,Olver1,Olver2,Stephani}
\def\sderef{Amir,Arnold,Evans,Fre,Ikeda,Kampen,Oksendal,Stroock}
\def\kozref{Koz1,Koz2,Koz3,Koz2018,Koz18a,Koz18b}
\def\symmstochref{GRQ1,GRQ2,Unal,Koz1,Koz2,Koz3,Koz2018,Koz18a,Koz18b,Glog,GL1,GL2,GS17,GGPR,GSW,GOU,GS20,GKS}

\begin{document}

\renewcommand{\evenhead}{ {\LARGE\textcolor{blue!10!black!40!green}{{\sf \ \ \ ]ocnmp[}}}\strut\hfill G. Gaeta \& M.A. Rodr\'iguez}
\renewcommand{\oddhead}{ {\LARGE\textcolor{blue!10!black!40!green}{{\sf ]ocnmp[}}}\ \ \ \ \   Integrable Ito equations with multiple noises}

\thispagestyle{empty}

\newcommand{\FistPageHead}[3]{
\begin{flushleft}
\raisebox{8mm}[0pt][0pt]
{\footnotesize \sf
\parbox{150mm}{{Open Communications in Nonlinear Mathematical Physics}\ \  \ {\LARGE\textcolor{blue!10!black!40!green}{]ocnmp[}}
\ \ Vol.2 (2022) pp
#2\hfill {\sc #3}}}\vspace{-13mm}
\end{flushleft}}

\FistPageHead{1}{\pageref{firstpage}--\pageref{lastpage}}{ \ \ \ }

\strut\hfill

\strut\hfill

\copyrightnote{The author(s). Distributed under a Creative Commons Attribution 4.0 International License}

\Name{Integrable autonomous scalar Ito equations with multiple noise sources}

\Author{Giuseppe Gaeta$^{\,1,\,2}$ and Miguel A. Rodr\'iguez$^{\,3}$}

\Address{$^{1}$  Dipartimento di Matematica, Universit\`a degli Studi di
Milano,via Saldini 50, 20133 Milano (Italy) \\[2mm]
$^{2}$ SMRI, 00058 Santa Marinella (Italy)\\[2mm]
$^{3}$ Departamento de F\'{\i}sica Te\'orica,
Universidad Complutense de Madrid, Pza.~de las Ciencias 1, 28040 Madrid (Spain)}

\Date{Received November 8, 2022; Accepted November 22, 2022}

\setcounter{equation}{0}

\begin{abstract}
\noindent The classification of scalar Ito equations with a single noise source which admit a so called standard symmetry and hence are -- by the Kozlov construction -- integrable is by now complete. In this paper we study the situation, occurring in physical as well as biological applications, where there are two independent noise sources. We determine all such autonomous Ito equations admitting a standard symmetry; we then integrate them by means of the Kozlov construction. We also consider the case of three or more independent noises, showing no standard symmetry is present.

\end{abstract}

\label{firstpage}


{

\tableofcontents
}

\section{Introduction}
\label{sec:intro}

Stochastic differential equations \cite{\sderef} are ubiquitous in Physics. These can be met in a variety of forms, but here we focus on those written in Ito form (for which a sound complete mathematical theory exists \cite{Ikeda}), i.e. as
\beql{eq:Itogen} d x^i \ = \ f^i (\xb , t) \ d t \ + \ \sum_{k=1}^\ell \s^i_{\ k}  (\xb , t) \ d w^k \ \ \ \ \ (i=1,...,n) \ , \eeq
where the $w^k = w^k (t)$ are independent Wiener processes.

Such equations can in general only be studied {\it per se} (that is, apart from the study of the associated diffusion -- Fokker-Planck or Kolomogorov -- equation) numerically, or at best qualitatively. But in a very small number of cases they are \emph{integrable}. In practice, this means that they can be set, through a change of variables $y = \Phi (x,t;w)$, in the form
\beql{eq:Itoint} d y^i \ = \ F^i (t) \ d t \ + \ \sum_{k=1}^\ell S^i_{\ k} (t) \ d w^k \ ; \eeq
equations in this form are then immediately integrated as
\beql{eq:introsimp} y^i (t) \ = \ y^i (0) \ + \ \int_0^t F^i (\tau) \ d \tau \ + \ \sum_{k=1}^\ell \int_0^t S^i_{\ k} (\tau) \ d w^k (\tau ) \ ; \eeq
inverting the change of variables allows to obtain $x(t) = \Phi^{-1} [ y(t)]$.

Note that contrary to deterministic integrable systems, in this case integrability does not entails that we are able to predict the state of the system at time $t >0$ once we know its initial state at $t=0$. In fact, the state $x(t)$ also depends on the \emph{realization} of the Wiener processes $w^k (\tau )$; thus prediction of $x(t)$ requires to know $x(0)$ \emph{and} the realizations of the Wiener processes. On the other hand, when one is dealing with these equations in the description of a physical (or biological) phenomenon, it is possible to use the explicit expression for the solution to test \emph{a posteriori} -- thus, knowing what the realization of the Wiener processes in the particular run of the experiment has been -- that the physical system is indeed described by the integrable equation.\footnote{For example, it is known that for several species the birth rate may depend on the temperature, and in some cases even the sex ratio among newborn is a function of the temperature \cite{Murray}; the temperature itself is a random variable, and in the case of an equation describing birth rate as a function of the temperature one can study \emph{a posteriori}, once the temperature fluctuations are known, the correspondence between these and fluctuations in the birth rate. Similarly, one can consider experiments where particles -- photons, electrons, etc -- are emitted by some electricity-driven apparatus, but random fluctuations in the electric current are present and taken into consideration; once this have been measured in a real run, correlation among them and those in the flow of emitted particles can be studied, and the model validated.}

In the case of systems, it may happen there one can eliminate the dependence on, say, $y^1,...,y^k$ in the drift and noise coefficients of the transformed system \eqref{eq:introsimp}; in this case we are effectively reduced to an $n-k$-dimensional system. That is, if $\{y^{k+1} (t),...,y^n (t) \}$ are determined, then the equations for $y^1,...,y^k$ are explicitly integrable \emph{reconstruction equations}. In this sense we speak of \emph{reduction}, or also \emph{partial integrability}, as in the case of deterministic equations \cite{\symmref}.\footnote{In the following we will mainly focus on scalar equations, for which the notion of partial integrability plays no role, but this would be relevant in higher-dimensional extensions of our work; see also the discussion in Section 5 of \cite{GL2} (Section 5 therein).}

It turns out that, similarly to what happens in the deterministic case \cite{\symmref}, integrability of Ito equations is strongly related to their \emph{symmetry properties} \cite{\symmstochref}. In fact, once a symmetry of \eqref{eq:Itogen} of a suitable form (see below for details) is known, there is a standard way of constructing the integrating change of variables, known as the \emph{Kozlov substitution} (so named after Roman Kozlov, who first devised it  \cite{\kozref}).

It should be noted that a somewhat weaker form of integrability also exists; that is, in some case -- depending on the functional form of the symmetry -- one will not be able to map the equation under study into \eqref{eq:Itoint}, but only into an equation of the form
\beql{eq:ItoWint} d y^i \ = \ F^i(t,\wb) \ dt \ + \ \sum_{k=1}^\ell S^i_{\ k} (t,\wb) \ dw^k \ . \eeq
This can still be formally integrated as
\beql{eq:ItoWint2} y^i (t) \ = \ y^i (t_0) \ + \ \int_{t_0}^t F^i [\tau , \wb (\tau) ] \ d \tau \ + \ \sum_{k=1}^\ell \int_{t_0}^t S^i_{\ k} [ \tau , \wb (\tau) ] \ d w^k (\tau) \ . \eeq

The equations \eqref{eq:Itogen} are rather general. In many cases, one is interested in \emph{autonomous} equations, and quite often \emph{scalar} ones (note that we are considering a scalar random variable $x(t)$, but allow this to be driven by several Wiener processes $w^k (t)$):
\beql{eq:Itot} d x \ = \ f(x) \ d t \ + \ \sum_{k=1}^\ell \s_k (x) \ d w^k \ . \eeq
This is the class of equations we consider in the present paper.

Moreover, albeit all forms of the noise terms $\s^i_{\ k} (\xb ,t)$ or $\s^i_{\ k} (\xb)$ or $\s_k (x,t)$ are in principles allowed, in most physical applications one is concerned with special forms of these, which we write in formulas (with $s$ a constant in each case) referring to \eqref{eq:Itot} for ease of notation:
\begin{itemize}
\item additive noise: $\s (x) = s$;
\item multiplicative noise: $\s (x) = s x$;
\item Poisson noise\footnote{Also known as fluctuational noise, RMS noise, demographical noise (in population dynamics), shot noise (in signal transmission theory), Schottky noise (in solid state Physics).}: $\s (x) = s \sqrt{x}$.
\item More generally, one can consider simple noise: $\s (x) = s x^k$ ($k$ a constant, say with $k \not= 0,1/2,1$ as these correspond to the previous special cases).
\end{itemize}

\noindent Later on, we will focus specially on these forms of noise.

We want to stress that a single system can be subject to different sources of noise: for example in Population Dynamics one will have fluctuations due to fluctuating environmental parameters (in this case one speaks in that context of \emph{environmental noise}, which is actually multiplicative noise in mathematical terms) and fluctuations due to intrinsic fluctuations in the birth and death processes, and these lead to Poisson noise (known in that context as \emph{demographical noise}).

Similarly, in situations where one is counting particles (electrons, photons, neutrons, quasi-particles, etc) reaching a detector, there will be a Poisson noise due to intrinsic fluctuations in the flow, but there can as well be multiplicative noise due to fluctuations in the operating conditions of the apparatus producing the particles or quasi-particles.

In such cases, where different sources of noise are present, such sources are in general \emph{not} correlated and the corresponding Wiener processes in the Ito equation should be seen as \emph{independent} ones. This lead us to the form \eqref{eq:Itogen} seen above; or to \eqref{eq:Itot} if we have a single variable $x(t)$.

In the simplest setting, one has a single variable $x$ and a single driving Wiener process $w(t)$.

In this context, it has been recently shown that the so called stochastic \emph{reaction equation} of chemical reaction theory (also known as \emph{logistic equation} in Population Dynamics) \emph{with multiplicative noise}, i.e. the (scalar, autonomous) Ito equation
\beql{eq:log} d x \ = \ A \ x \ (1 - x ) \ d t \ + \ s \ x \ d w \ , \eeq
is integrable. For this result, and the actual integration, see \cite{Glog}.
Search for more general integrable scalar stochastic differential equations with \emph{multiplicative} noise ensued, and these have been classified in \cite{GS20}.

Another classification, devoted to integrable scalar equations with general \emph{simple noise} (that is, for noise term of the form $s x^k$, with $k \not= 1$ -- this case being covered by the work mentioned above, and also being degenerate in several ways) was given recently \cite{GR22}. This classification covers, as said above, any simple noise and hence also the special cases of additive noise and of Poisson noise. (This classification actually covers also the case of a general -- i.e. non-simple -- noise, but in a rather implicit manner.)

In this sense, at present we know all we would like to know about integrable scalar autonomous Ito equations with a \emph{single} noise source.

The goal of this study is to investigate the situation for systems described by a single random variable $x(t)$ subject to \emph{multiple} noise sources. In particular, we will consider the case where the noise is a combination of two of the special types of noise listed above: additive, multiplicative and Poisson; or general simple noise as well.

As stressed above, beside restricting to the case of scalar equations, we will focus on the physically relevant case of \emph{autonomous} Ito equations. Thus our subject of study will be equations of the form
\beql{eq:Ito} d x \ = \ f(x) \ d t \ + \ \sum_{k=1}^\ell \s_k (x) \ d w^k \ , \eeq actually with $\ell = 2$.

We will from now on routinely use the Einstein summation convention, albeit in some cases we may prefer to write again explicitly the summations for greater clarity. Moreover we will write $\pa_x$ for $\pa / \pa x$ and the like. In later sections, when we deal with only one random variable $x$ but several driving Wiener processes $w^i$, we will sometimes write $\pa_k$ for $\pa / \pa w^k$.

\addcontentsline{toc}{subsection}{Plan of the paper}

\subsection*{Plan of the paper}

The \emph{plan of the paper} is as follows. In Section \ref{sec:sym} we recall some facts about symmetry and integrability -- and the interrelations between the two -- of Ito equations. This also identifies the class of relevant symmetries in the present context, and we will then restrict to consider only these, dibbed simply as ``symmetries'' in this description of the plan of the paper. We pass then, in Section \ref{sec:symmtwo}, to discuss the case where there are \emph{two} noise sources (driving Wiener processes); this is the core of the paper, and albeit we focus on the (physically more relevant, as discussed above in this Introduction) case of two noises, the methods used here are promptly extended to the case of multiple noise sources. In Section \ref{sec:CC} we find a \emph{compatibility condition} which must be satisfied by the two noise terms; this leads to a general result: if we are considering \emph{simple} noises, then for a symmetry to exist one of the two noises must be a multiplicative one. We pass then to consider the most relevant -- in terms of Physics applications -- combination of noise types (including combinations for which the aforementioned general results implies no symmetry can be present, checking this result by a different approach): additive and multiplicative, additive and Poisson, Poisson and multiplicative, multiplicative and general simple noise. For each of these cases, we classify the symmetric equations -- that is, the allowed drift term and possibly relations between the constant appearing in the noise coefficients of the given functional type corresponding to an equation admitting symmetries -- and for these we also determine the symmetries. In the following Section \ref{sec:inttwo} we will use these symmetries to integrate, via the Kozlov substitution, the symmetric equations. It is well known that when dealing with an Ito equation a standard substitution \cite{Koz1,Koz2,Koz3} allows to set \emph{one} of the noise terms to unity. One could use this approach in studying equations with two noise terms, thus reducing these to a kind of ``standard form''; in Section \ref{sec:oneconst} we use this approach to double check our results described before. Section \ref{sec:NNoises} briefly discusses the case of a scalar (autonomous) Ito equation with $N$ noises; it turns out that there is no symmetry for $N > 2$, so that our classification is complete. Finally, in Section \ref{sec:discussion} we discuss our findings and draw our conclusions.

The paper is completed by three Appendices. In Appendix \ref{app:FODE} we recall, following also our recent paper \cite{GR22} (but extending the scope of the discussion given there to the case of multiple noises), how the second order symmetry determining equations can be cast in first order form by considering the Stratonovich drift for the Ito equation. Appendix \ref{app:NOW} is devoted to justifying the restriction, in the main body of the paper, to \emph{standard} symmetries (see Sect.\ref{sec:sym}), excluding so called \emph{W-symmetries} \cite{GSW}. In Appendix \ref{app:3N} we provide the detailed computations showing that no standard symmetry can arise for autonomous scalar equations with three independent noises, so that our classification is actually complete -- not only for two noises but for multiple independent noises in general.


\section{Symmetry and integrability of scalar Ito equations. \\ General aspects}
\label{sec:sym}

In this Section we review some well know facts about symmetry and integrability of Ito equations, applying these to the case of interest here.

\subsection{Symmetry}
\label{sec:symsym}

A time-preserving symmetry of a general scalar Ito equation with possibly multiple sources of noise\footnote{The restriction to scalar equations is inessential from the mathematical point of view, but handy here in that: (i) this is the case we will treat; (ii) it allows to provide slightly simpler formulas. See e.g. \cite{GGPR} for the fully general case.} is a vector field
\beql{eq:Ygen} Y \ = \ \vphi (x,t;\wb) \, \pa_x \ + \ (R \, \wb)^i \, \pa_{w^i}  \eeq
which satisfies the \emph{determining equations}
\begin{eqnarray}
& & \vphi_t \ + \ f \, \vphi_x \ - \ \vphi \, f_x \ + \ \frac12 \, \Delta (\vphi ) \ = \ 0 \ , \label{eq:deteq1gen} \\
& & \vphi_{w^i} \ + \ \s^i \, \vphi_x \ - \ \vphi \, \s^i_x \ - \ (R^i_{\ k} \, \s^k) \ = \ 0 \ \ \ (i = 1 , ... , \ell) ; \label{eq:deteq2gen} \end{eqnarray}
here and in the following, $\Delta$ is the \emph{Ito Laplacian}, which in this simple scalar case reads
\beql{eq:Delta} \Delta (\phi) \ := \ \sum_{k=1}^\ell \phi_{w^k w^k} \ + \ 2 \, \sum_{k=1}^\ell \s_k \, \phi_{xw^k} \ + \ \( \sum_{k=1}^\ell \s_k^2 \) \, \phi_{xx} \ . \eeq

Moreover, $R$ is a constant matrix belonging to the Lie algebra of the conformal linear group in $\ell$ dimensions.
When $R=0$, we speak of \emph{standard} symmetries, while for $R \not= 0$ we have proper \emph{W-symmetries}\footnote{In the usual nomenclature, standard symmetries are a special case of W-symmetries; this is why when $R\not= 0$, hence we have a symmetry which is \emph{not} a standard one, we speak of \emph{proper} W-symmetries.}. Moreover we distinguish two types of standard symmetries:  when $\vphi = \vphi (x,t)$ we speak of a \emph{deterministic} standard symmetry, while if $\vphi$ depends on (at least some of) the $w^i$, we have a \emph{random} symmetry.

It should be noted that standard symmetries lead, through the Kozlov substitution, to (full or partial) integration of the equation via a change of variables which maps the equation at hand into a new (fully or partially integrable; see the Introduction) stochastic equation, of Ito type in the case of deterministic standard symmetries and of a related type in the case of random standard symmetries. In the case of W-symmetries the situation is less well defined, see \cite{GSW,GR22} for details; actually, we will show in Appendix \ref{app:NOW} that W-symmetries can not be used for integration of Ito equations though a change of variables; so they will not be considered by our study.

In the present setting, we should solve \eqref{eq:deteq1gen}, \eqref{eq:deteq2gen} for given $\s_k$ as an equation for both the drift $f$ and $\{ \vphi , R \}$ (the coefficients of the symmetry vector field).

Note that our system is a system of \emph{linear} equations for $\vphi$, and for $R=0$ this is a \emph{homogeneous} system; thus -- at least in this $R=0$ case, i.e. standard symmetries, which is the one we will deal with -- $\vphi$ can always be multiplied by a constant; in our forthcoming concrete computations we will set this to unity, i.e. disregard this unessential multiplicative constant.

\subsection{Integrability}
\label{sec:symint}

If a standard symmetry (exists and) has been determined, we consider the so called \emph{Kozlov transformation}, i.e. define the function
\beq \Phi(x,t;\wb) \ = \ \int \frac{1}{\vphi (x,t;\wb)} \ d x \ , \eeq
and operate the change of variables
\beq y \ = \ \Phi (x,t;\wb) \ . \eeq

It is well known that this explicitly integrates our equation \cite{Koz2}; we will however provide a brief description of how the proof goes, for the sake of completeness. (We refer to our recent paper \cite{GR22} for details).

We will use the shorthand notation
$$ \vphi_k \ := \ \frac{\pa \vphi}{\pa w^k} \ . $$
Applying Ito formula, the new random variable $y$ satisfies the Ito equation
\beq d y \ = \ \frac{\pa \Phi}{\pa x} \, d x \ + \ \frac{\pa \Phi}{\pa t} \, d t \ + \ \frac{\pa\Phi}{\pa w^k} \, d w^k \ + \ \frac12 \, \Delta (\Phi) \, d t \ . \eeq

We should now substitute for $dx$ according to \eqref{eq:Ito}. In this way we obtain an equation of the form
\beq d y \ = \ F \, dt \ + \ S_k \, dw^k \ , \eeq
and by explicit computations we have
\begin{eqnarray*}
F &=& \frac{f}{\vphi} \ - \ \sum_k \( \frac{\s_k \, \vphi_k}{\vphi^2} \ +  \ \frac{\s_k^2 \vphi_x}{2 \, \vphi^2} \) \ - \ \frac12 \, \sum_k \( \int \frac{\vphi \, \vphi_{kk} \, - \, 2 \, \vphi_k^2}{\vphi^3} \ d x \) \ - \ \int \frac{\vphi_t}{\vphi^2} \ d x \ ; \\
S_k &=& \frac{\s_k}{\vphi} \ - \ \int \frac{\vphi_k}{\vphi^2} \ d x \ . \end{eqnarray*}
Note that in these formulas the r.h.s. is written as a function of $x$; this should be thought as $x = \Phi^{-1} (y)$, but as we will see in a moment we do not actually need to perform the substitution in order to show that $F$ and the $S_k$ only depend on $t$, and possibly on the $w^k$, but not on $y$.

In order to check that $F$ and $S$ do not depend on $y$, we should apply the $y$ derivative on these expressions; that is, in view of the change of variables we have performed (in which $t$ and $w_k$ are unaffected), we should consider the operator $$ \frac{\pa}{\pa y} \ = \ \( \frac{\pa x}{\pa y} \) \ \frac{\pa}{\pa x} \ = \ \vphi \ \frac{\pa}{\pa x} \ . $$ Thus, if $(\pa F /\pa x) = 0$, $(\pa S_k / \pa x) = 0$, we are automatically guaranteed to have $(\pa F / \pa y ) = 0$, $(\pa S_k / \pa y) = 0 $ as well.

Actually, if we consider the determining equations and substitute for $\vphi_t $ and for $\vphi_k$ (and its differential consequences, i.e. -- with an obvious notation -- for $\vphi_{kk}$ and $\vphi_{xk}$) according to these, we readily obtain that
\beq \frac{\pa F}{\pa x} \ = \ 0 \ ; \ \ \frac{\pa S_k}{\pa x} \ = \ 0 \ . \eeq
In fact, we immediately get
$$ \frac{\pa S_k}{\pa x} \ = \ \frac{1}{\vphi^2} \ \[ \vphi_k \ + \ \s_k \, \vphi_x \ - \ \vphi \, (\pa \s_k / \pa x) \] \ ; $$
if $\vphi$ satisfies \eqref{eq:deteq2gen} (recalling we are assuming to have a standard symmetry, i.e. $R=0$) it is immediate to see this vanishes.
The computation for $F$ is more involved and will not be reported here; it makes use of \eqref{eq:deteq1gen} and \eqref{eq:deteq2gen} together with its differential consequences, and is completely analogous to the one for $S_k$; it is given in full detail in the  Appendix A to \cite{GR22}.

It should be stressed that proceeding in the same way for W-symmetries  we find that
$$ \frac{\pa S_k}{\pa x} \ = \  \frac{R_{k \ell} \s_\ell}{\vphi^2} \ , $$
which is nonzero for $R \not= 0$, i.e. for proper W-symmetries.

The reader has surely noted that we proved only that $F$ and $S$ are independent of $x$, and hence of $y$; \emph{not} that they are also independent of the $w^k$. In fact, in general they \emph{do} depend on these variables. We are guaranteed this is not the case, of course, when $\vphi$ itself does not depend on the $w^k$, $\vphi = \vphi (x,t)$; that is, when we have a so called \emph{deterministic} standard symmetry.

In this case computations are simpler: we do have
\begin{eqnarray}
F &=& \frac{f}{\vphi} \ - \ \frac{\s^2 \ \vphi_x}{2 \, \vphi^2} \ - \ \int \frac{\vphi_t}{\vphi} \, dx \ , \\
S_k &=& \frac{ \s_k}{\vphi} \ , \end{eqnarray}
and checking the $x$-independence of these once we assume $\vphi$ is a solution to the determining equations (with $R=0$) is immediate.

We refer to \cite{Koz2,GL2} for detailed proofs, also dealing with more general cases than those considered in this paper.

It should be stressed again that when $F$ and the $S_k$ do actually depend on the $w^k$, the transformed equation is not in Ito form; but it is still integrable. In fact we do have the transformed equation in the form \eqref{eq:ItoWint}, which is integrated as \eqref{eq:ItoWint2}.

\section{Standard symmetries of scalar Ito equations with two Wiener processes}
\label{sec:symmtwo}

As recalled in the Introduction, the results available in the literature -- see in particular our recent paper \cite{GR22}, fully cover the case of a scalar Ito equation with a  \emph{single} noise term.

But a scalar Ito equation can have \emph{multiple} noise terms, and it is natural to discuss this class of equations as well. Actually, as discussed in the Introduction, in several physical or biological situations one has to consider a Ito equation with \emph{two} noise terms; this is of course also the simplest case of multiple noise terms.

We will thus from now on consider the framework of a single (autonomous) Ito equation for a scalar random variable $x$ having two independent driving Wiener processes\footnote{It is rather clear that in case of linear dependence -- which for two noises would simply mean they are one multiple of the other -- we can reduce to consider a smaller number of noise sources. We will thus always assume, both here and in considering the $N$ noises case in Sect. \ref{sec:NNoises} below, that such dependencies -- and the ensuing reduction -- have been taken into account.}. This would be written in the form \eqref{eq:Ito} with $\ell = 2$, but for notational simplicity we will set
$$ \s_1 (x) \ = \ \s (x) \ , \ \ \s_2 (x) \ = \ \vr (x) \ ; \ \ w_1 \ = \ w \ , \ \ w_2 \ = \ z \ . $$ Thus we write this class of equations explicitly as
\beql{eq:Itotwo} d x \ = \ f(x) \, d t \ + \ \s (x) \, d w \ + \ \vr (x) \, d z \ . \eeq
In this case the determining equations consist of a system of three equations:
\begin{eqnarray}
& & \vphi_t \ + \ f \, \vphi_x \ - \ \vphi \, f_x \ + \ \frac12 \, \Delta (\vphi ) \ = \ 0 \ , \label{eq:deteq1nso} \\
& & \vphi_w \ + \ \s \, \vphi_x \ - \ \vphi \, \s_x \ - \ r_{11} \, \s \ - \ r_{12} \, \vr \ = \ 0 \ , \label{eq:deteq2ar} \\
& & \vphi_z \ + \ \vr \, \vphi_x \ - \ \vphi \, \vr_x \ - \ r_{21} \, \s \ - \ r_{22}  \, \vr \ = \ 0 \ ; \label{eq:deteq2br}
\end{eqnarray}
the Ito Laplacian is now defined considering we have two (independent) driving Wiener processes, which gives in concrete terms
\beql{eq:Deltan} \Delta (\phi) \ := \ \phi_{ww} \ + \ \phi_{zz} \ + \ 2 \, \( \s \, \phi_{xw} \ + \ \vr \, \phi_{xz} \) \ + \ \( \s^2 \ + \ \vr^2 \) \, \phi_{xx} \ . \eeq
For later reference we will group the $r_{ij}$ coefficients in a matrix
$$ R \ = \ \begin{pmatrix} r_{11} & r_{12} \\ r_{21} & r_{22} \end{pmatrix} \ . $$

As in the case of a single noise \cite{GR22}, this system can be replaced by a first order system, in which the place of \eqref{eq:deteq1nso} is taken by
\beql{eq:deteq1nfor} \vphi_t \ + \ b \, \vphi_x \ - \ b_x \, \vphi \ + \ r_{11} \, \s \, \s_x \ + \ r_{22} \, \vr \, \vr_x \ + \ \frac12 \ \(  r_{12} \, + \, r_{21} \) \ \( \s \, \vr_x \ + \ \s_x \, \vr \) = \ 0  \ , \eeq
and \eqref{eq:deteq2ar}, \eqref{eq:deteq2br} remain unchanged (see Appendix \ref{app:FODE} for a proof). Note that, as already remarked, the $b$ thus defined implicitly, and explicitly given by
\beql{eq:btwodim} b \ = \ f \ - \ \frac12 \ \( \s \, \s_x \ + \ \vr \, \vr_x \) \ , \eeq is the \emph{Stratonovich drift} for our Ito equation \cite{\sderef}.

The symmetries (corresponding to solutions of the system above) will be of the form
\beql{eq:Ygennr} Y \ = \ \vphi (x,t;w,z) \, \pa_x \ + \ \( r_{11} \, w \ + \ r_{12} \, z \) \, \pa_w \ + \ \( r_{21} \, w \ + \ r_{22} \, z \) \, \pa_z \ . \eeq

We will from now on, in this Section and the following ones, only deal with the case
\beq R \ = \ 0 \ ; \eeq
that is, we will only consider \emph{standard} symmetries. The reason to exclude proper W-symmetries is, as mentioned above, that they do not lead to integration of the Ito equation through a change of variables; this matter will be discussed in detail in Appendix \ref{app:NOW} below.

In this way, eqs. \eqref{eq:deteq1nfor}, \eqref{eq:deteq2ar} and \eqref{eq:deteq2br} reduce to
\begin{eqnarray}
& & \vphi_t \ + \ b \, \vphi_x \ - \ \vphi \, b_x \ = \ 0  \ ; \label{eq:deteq1n} \\
& & \vphi_w \ + \ \s \, \vphi_x \ - \ \vphi \, \s_x \ = \ 0 \ , \label{eq:deteq2a} \\
& & \vphi_z \ + \ \vr \, \vphi_x \ - \ \vphi \, \vr_x  \ = \ 0 \ , \label{eq:deteq2b} \end{eqnarray}
while \eqref{eq:deteq1nso} is of course unchanged. Moreover, the vector field \eqref{eq:Ygennr} is just
$ Y  = \vphi (x,t;w,z) \, \pa_x$.

When tackling the system \eqref{eq:deteq1n}, \eqref{eq:deteq2a}, \eqref{eq:deteq2b}, our general approach will consist in first solving the equations which are independent of $f$, i.e. \eqref{eq:deteq2a} and \eqref{eq:deteq2b}; this determines a general class of functions $\vphi$. We will then turn our attention to \eqref{eq:deteq1n}, where we have to determine both $f$ and the exact function $\vphi$ within the class identified before; in general, it will be convenient to consider differential consequences of \eqref{eq:deteq1n} providing some separation of variables (see below for details).

In this section we will just identify symmetrical equations and determine their symmetry; in the next section we will use the latter to integrate the symmetrical equations.

\subsection{Compatibility condition}
\label{sec:CC}

The equations \eqref{eq:deteq2a} and \eqref{eq:deteq2b} admit a common solution only if a certain compatibility condition between the noise terms $\s (x)$ and $\vr (x)$ is satisfied. Note that we should assume these to be functionally independent, or the two equations would be equivalent. A special situation arises in the case of simple noises.

\medskip\noindent
{\bf Lemma 1.} {\it Let $\s (x)$ and $\vr (x)$ be functionally independent. The equations \eqref{eq:deteq2a} and \eqref{eq:deteq2b} admit a common solution $\vphi$ if and only if the functions
\begin{eqnarray}
J(x) &:=& \frac{ \s \ \( \s \, \vr_{xx} \ - \ \s_{xx} \, \vr \) \ - \ \s_x \ \( \s \, \vr_x \ - \ \s_x \, \vr \)}{\s \, \vr_x \ - \ \s_x \, \vr } \ , \label{eq:J} \\
K(x) &=& \frac{ \vr \ \( \s \, \vr_{xx} \ - \ \s_{xx} \, \vr \) \ - \ \vr_x \ \( \s \, \vr_x \ - \ \s_x \, \vr \)}{\s \, \vr_x \ - \ \s_x \, \vr } \ , \label{eq:K}  \end{eqnarray}
are actually constant.}

\medskip\noindent
{\bf Proof.} Taking the $z$ derivative of \eqref{eq:deteq2a} and the $w$ derivative of \eqref{eq:deteq2b}, we obtain
\begin{eqnarray}
\vphi_{wz} &=& \s_{x} \ \vphi_z  \ - \ \s \ \vphi_{xz} \ , \label{eq:pL1a} \\
\vphi_{zw} &=& \vr_{x} \ \vphi_w  \ - \ \vr \ \vphi_{xw}  \ ; \label{eq:pL1b} \end{eqnarray}
the last term in each equation is obtained differentiating \eqref{eq:deteq2a} and \eqref{eq:deteq2b} w.r.t. $x$, which gives
\begin{eqnarray*}
\vphi_{xw} &=& \vphi \, \s_{xx} \ - \s \, \vphi_{xx} \ , \\
\vphi_{xz} &=& \vphi \, \vr_{xx} \ - \ \vr \, \vphi_{xx} \ . \end{eqnarray*}
Taking these and \eqref{eq:deteq2a}, \eqref{eq:deteq2b} into account, the difference of \eqref{eq:pL1a} and \eqref{eq:pL1b} yields
\beq \( \s \ \vr_x \ - \ \s_x \ \vr \) \ \vphi_x \ = \ \( \s \ \vr_{xx} \ - \ \s_{xx} \ \vr \) \ \vphi \ . \eeq
Writing for short
\beql{eq:LA} \La (x) \ := \ \s \ \vr_x \ - \ \s_x \ \vr \ , \eeq
this equation yields
\beql{eq:phicc} \vphi (x,t;w,z) \ = \ \La (x) \ h(t;z,w)  \ . \eeq
Note that $\La (x) \not= 0$ unless $\s$ and $\vr$ are functionally dependent, which we exclude.

The functional form \eqref{eq:phicc} corresponds to a compatibility condition between \eqref{eq:deteq2a} and \eqref{eq:deteq2b}. Plugging this expression for $\vphi$ into \eqref{eq:deteq2a} and \eqref{eq:deteq2b}, we get
$$
\La \ h_w \ + \ \( \s \, \La_x \ - \ \s_x \, \La \) \ h \ = \ 0 \ , \ \ \ \
\La \ h_z \ + \ \( \vr \, \La_x \ - \ \vr_x \, \La \) \ h \ = \ 0 \ ; $$  these are also rewritten as
\begin{eqnarray}
\frac{h_w}{h} &=& \frac{\La \, \s_x \ - \ \La_x \, \s}{\La} \ := \ - \, J \ , \nonumber \\
\frac{h_z}{h} &=& \frac{\La \, \vr_x \ - \ \La_x \, \vr}{\La} \ :=  - \, K \ . \end{eqnarray}
As $h = h (t;z,w)$, the l.h.s. does not depend on $x$, and hence $J$ and $K$ defined above must also not depend on $x$:
\beql{eq:JKconst} \frac{\pa J}{\pa x} \ = \ 0 \ = \ \frac{\pa K}{\pa x} \ . \eeq

If we express $\La$ in terms of $\s$ and $\vr$, see \eqref{eq:LA}, we have precisely the expressions \eqref{eq:J} and \eqref{eq:K} given in the statement. \EOP

\medskip\noindent
{\bf Corollary 1.} {\it If $\s$ and $\vr$ are simple noises, i.e. (with $s,r$ nonzero real constants) \beql{eq:S2N} \s (x) \ = \ s \ x^\ell \ , \ \ \ \vr (x) \ = \ r \ x^m \ , \eeq then the equations \eqref{eq:deteq2a} and \eqref{eq:deteq2b} admit a common solution if and only if either $\ell = 1$ or $m=1$, i.e. if and only if one of the two noises is a multiplicative noise.}

\medskip\noindent
{\bf Proof.} This follows by the definition of $\La$. In the case \eqref{eq:S2N} we have $$ \La \ = \ r \ s \ (m - \ell) \ x^{\ell + m - 1} $$ and hence
\beq J \ = \ (m \, - \, 1 ) \ s \ x^{\ell-1} \ \ , \ \ \ \ K \ = \ (\ell \, - \, 1) \ r \ x^{m-1} \ . \eeq
These are constant only when either the exponent of $x$, or the coefficient, is zero. The statement follows at once. \EOP
\bigskip

It is a trivial remark that this approach, through compatibility conditions, could as well be pursued with an arbitrary number $N$ of independent noise sources. We will only consider in detail the combination of two noises, as anticipated. The case of $N$ noises is then discussed in Section \ref{sec:NNoises}; we anticipate that this discussion will show that only the two-noises case allows nontrivial symmetry properties.

\subsection{Equations with additive and multiplicative noise}

We start by considering the case where the equation has additive and multiplicative noise, i.e. we deal with
\beql{eq:Itoam} d x \ = \ f(x) \ dt \ + \ s_1 \ d w \ + \ s_2 \, x \ d z \ . \eeq

\medskip\noindent
{\bf Lemma 2.} {\it An equation of the form \eqref{eq:Itoam} admits a standard symmetry if and only if the drift term $f(x)$ is of the form \beql{eq:fam} f(x) \ = \ \a \ + \ \b \ x \ , \eeq with $\a$ and $\b$ real constants.
In this case, the symmetry is identified by
\beql{eq:phiam} \vphi (x,t;w,z) \ = \ \exp \[ s_2 \, z \ + \ (\b \, - \, s_2^2 / 2) \ t \] \ . \eeq
}

\medskip\noindent
{\bf Proof.}
In this case the equations \eqref{eq:deteq2a} and \eqref{eq:deteq2b} read
\begin{eqnarray*}
& & \vphi_w \ + \ s_1 \ \vphi_x \ = \ 0 \ , \\
& & \vphi_z \ + \ s_2 \, x \ \vphi_x \ = \ s_2 \ \vphi \ . \end{eqnarray*}
These are easily solved by characteristics, yielding
\beq \vphi (x,t;w,z) \ = \ \exp [ s_2 \ z] \ \eta(t) \ . \eeq

We have now to deal with \eqref{eq:deteq1n}, which now reads simply
\beql{eq:eq1am} \frac12 \ \exp[ s_2 \, z] \ \[ 2 \, \eta' (t) \ + \ \( s_2^2 \ - \ 2 \, f' (x) \) \] \ = \ 0 \ ; \eeq
dropping the overall factor $(1/2)\exp[ s_2 z]$ and differentiating w.r.t. $x$ we get that $\eta \not= 0$ requires
$f(x)$ to be precisely of the form \eqref{eq:fam}.
The equation \eqref{eq:eq1am} reads now
\beq 2 \ \eta' \ + \ (s_2^2 \ - \ 2 \, \b ) \ \eta \ = \ 0 \ , \eeq hence (disregarding as usual the multiplicative constant) we get
\beq \eta (t) \ = \ \exp \[ (\b \ - \ s_2^2/2) \ t \] \ . \eeq
This completes the proof. \EOP
\bigskip

Note that the symmetry thus determined is a random standard one; as discussed in Section \ref{sec:symint}, this will lead to an integrable stochastic equation which is \emph{not} in Ito form.

\subsection{Equations with additive and Poisson noise}

We know that in this case no symmetry can be present, see the discussion in Section \ref{sec:CC}. We will however perform the direct computations to confirm our result.

Let us now consider autonomous scalar Ito equations depending on both additive and Poisson noise, i.e.
\beql{eq:Itoap} d x \ = \ f(x) \ d t \ + \ s_1 \, d w \ + \ s_2 \ \sqrt{x} \ d z \ . \eeq

\medskip\noindent
{\bf Lemma 3.} {\it An equation of the form \eqref{eq:Itoap} never admits a standard symmetry.}

\medskip\noindent{\bf Proof.}
In this case the determining equations \eqref{eq:deteq2a} and \eqref{eq:deteq2b} read respectively
\begin{eqnarray}
& & \vphi_w \ + \ s_1 \ \vphi_x \ = \ 0 \ , \label{eq:deteq2a_ap} \\
& & \vphi_z \ + \ s_2 \ \sqrt{x} \, \vphi_x \ = \ \frac{s_2}{2 \, \sqrt{x}} \ \vphi \ . \label{eq:deteq2b_ap} \end{eqnarray}
The first equation yields immediately
$$ \vphi (x,t;w,z) \ = \ \psi (t,\zeta,z) \ \ , \ \ \ \ \zeta := w - x/s_1 \ . $$
Plugging this into \eqref{eq:deteq2b_ap} we get the equation
$$ \psi_z \ - \ \frac{s_2}{s_1} \ \sqrt{x} \ \psi_\zeta \ = \ \frac{s_2}{2 \, \sqrt{x}} \ \psi \ . $$ Recalling $\psi = \psi (t,\zeta,z)$ we obtain that necessarily $\psi = 0$ and hence $\vphi = 0$ as well. \EOP

\bigskip
It should be noted that this also follows from our general result given in Lemma 1; we have performed this explicit computation as a confirmation of this, also considering that the one presently considered is a relevant combination of independent noises for the applications.

\subsection{Equations with Poisson and multiplicative noise}

We can now consider equations with both Poisson and multiplicative noise. These are
\beql{eq:Itopm} d x \ = \ f (x) \, dt \ + \ s_1 \, \sqrt{x} \, dw \ + \ s_2 \, x \, dz\ . \eeq

\medskip\noindent
{\bf Lemma 4.} {\it An equation of the form \eqref{eq:Itopm} admits a standard symmetry if and only if the drift term $f(x)$ is of the form \beql{eq:fpm} f(x) \ = \ (s_1^2 / 4)  \ + \ c_2 \ \sqrt{x} \ + \ c_3 \ x \ ; \eeq in this case, the symmetry is identified by
    \beql{eq:phipm} \vphi (x,t;w,z) \ = \ \sqrt{x} \ \exp \[ \frac14 \ \( 2 \, s_2 \, z \ + \ (2 \, c_3 \ - \ s_2^2 ) \ t \) \] \ . \eeq
}

\medskip\noindent
{\bf Proof.}
We will start once again by considering eqs. \eqref{eq:deteq2a} and \eqref{eq:deteq2b}. These now read
\begin{eqnarray}
& & \vphi_w \ + \ s_1 \ \sqrt{x} \ \vphi_x \ = \ \frac{s_1}{2 \sqrt{x}} \ \vphi \ , \label{eq:eq2a_pm} \\
& & \vphi_z \ + \ s_2 \ x \ \vphi_x \ = \ s_2 \ \vphi \ . \label{eq:eq2b_pm} \end{eqnarray}

We start by considering \eqref{eq:eq2b_pm}, which is solved by
$$ \vphi (x,t;w,z) \ = \ x \ \chi (t,w,\zeta ) \ \ , \ \ \ \ \zeta := z - \frac{\log(x)}{s_2} \ . $$
Plugging this into \eqref{eq:eq2a_pm} we get
$$ x \ \chi_w \ - \ \frac{s_1}{s_2} \ \sqrt{x} \ \chi_\zeta \ = \ \frac{s_1}{2} \ \sqrt{x} \ \chi \ . $$ This requires $\chi_w=0$, i.e.
$$ \chi (t,w,\zeta) \ = \ \psi (t,\zeta) \ . $$ The equation \eqref{eq:eq2a_pm} is thus reduced to
\beq \frac{s_1}{2 \, s_2} \ \sqrt{x} \ \( s_2 \ \psi \ - \ 2 \ \psi_\zeta \) \ = \ 0 \ , \eeq
with solution
\beq \psi (t,\zeta) \ = \ \exp[ (s_2/2) \ \zeta] \ \eta (t) \ . \eeq

We are now ready to tackle \eqref{eq:deteq1n}, which now reads -- factoring out an overall term $\exp[(s_2/2) z]/(8 \, \sqrt{x})$ -- as
\beql{eq:eq1pm}  8 \, x \, \eta' \ - \ \( s_1^2 \ - \ 2 \, s_2^2 \, x \ - \ 4 \, f(x) \ + \ 8 \, x \, f' (x) \) \ \eta \ = \ 0 \ . \eeq
Differentiating twice w.r.t. $x$ we have
$$ 4 \ \eta \ \[ 3 \, f'' (x) \ + \ 2 \, x \, f''' (x) \] \ = \ 0 \ . $$ As we want $\eta \not= 0$, this requires
$$ f(x) \ = \ c_1 \ + \ c_2 \ \sqrt{x} \ + \ c_3 \ x \ . $$

Having restricted the functional form of $f(x)$, we can go back to \eqref{eq:eq1pm}, which now reads
\beq \( s_1^2 \ - \ 4 \, c_1 \) \ \eta \ + \ x \ \[ \( 4 \, c_3 \ - \ 2 \, s_2^2 \) \ \eta \ - \ 8 \ \eta' \] \ = \ 0 \ . \eeq
The terms corresponding to different powers of $x$ must vanish separately, so we have
$$ c_1 \ = \ s_1^2 / 4 $$ from the terms of order zero, which sets $f(x)$ to be precisely of the form \eqref{eq:fpm}, while the term of order one in $x$ yields (disregarding the multiplicative constant)
$$ \eta (t) \ = \ \exp \[ \( \frac{c_3}{2} \ - \ \frac{s_2^2}{4} \) \ t \] \ . $$
This completes the proof. \EOP

\bigskip
In this case again we have a random standard symmetry, and as discussed in Section \ref{sec:symint} this will lead to an integrable equation which is not in Ito form.

\subsection{Equations with multiplicative and general simple noise}

We have considered above the combinations of noises which more often arise in physical application. More generally, we could consider the combination of two simple noises (leading to rather complex direct computations); but our Corollary 1 shows that one of these should be a multiplicative noise, or no symmetry can be present.

Thus we will now just consider, for the sake of completeness, the combination of a multiplicative and of a general simple noise, i.e.
$$ \s (x) \ = \ s_1 \ x^m \ , \ \ \ \vr (x) \ = \ s_2 \ x \ . $$
In other words, we will consider the Ito equations
\beql{eq:Itoms} d x \ = \ f(x) \, d t \ + \ s_1 \, x^m \, d w \ + \ s_2 \, x \, d z \ . \eeq
We stress that we should and will require
$$ m \ \not= \ 1 \ . $$
In the following computations, we will also assume $m \not=0$, $m \not= 1/2$, as these cases have already been considered (this will allow to avoid discussing certain degenerations in the forthcoming general formulas).

\medskip\noindent
{\bf Lemma 5.} {\it An equation of the form \eqref{eq:Itoms} admits a standard symmetry if and only if the drift term $f(x)$ is of the form \beql{eq:fms} f(x) \ = \ \frac12 \ \( (s_2^2 \, - \, 2 \, Q) \, x \ + \ m \, s_1^2 \, x^{2 m - 1} \) \ + \ C \, x^m \ , \eeq with $C$ and $Q$ real constants; in this case, the symmetry is identified by
\beql{eq:phims} \vphi (x,t;w,z) \ = \ x^m \ \exp \[ (m-1) \  (Q \, t \ - \ s_2 \, z ) \] \ . \eeq}

\medskip\noindent
{\bf Proof.}
We start as usual by considering eqs. \eqref{eq:deteq2a} and \eqref{eq:deteq2b}. These now read
\begin{eqnarray}
& & \vphi_w \ + \ s_1 \ x^m \ \vphi_x \ = \ s_1 \ m \ x^{m-1} \ \vphi \ , \label{eq:eq2a_ms} \\
& & \vphi_z \ + \ s_2 \ x \ \vphi_x \ = \ \ s_2 \ \vphi \ . \label{eq:eq2b_ms} \end{eqnarray}

The equation \eqref{eq:eq2b_ms} is solved by
$$ \vphi (x,t;w,z) \ = \ x \ \chi (t,\theta , w ) \ \ , \ \ \ \ \theta :=  z - \frac{\log(x)}{s_2} \ . $$
Plugging this into \eqref{eq:eq2a_ms} we get
$$ s_2 \, x \ \chi_w \ - \ s_1 \ x^m \ \chi_\theta \ = \ (m \, - \, 1) \, s_2 \, s_1 \, x^m \ \chi \ . $$ This requires $\chi_w =0$, i.e.
$$ \chi (t,\theta,w) \ = \ \psi (t,\theta) \ . $$ The equation \eqref{eq:eq2a_ms} is thus reduced, dropping an overall $s_1 x^m$ factor, to
\beq  s_2 \, (m-1) \ \psi \ + \ \psi_\theta  \ = \ 0 \ , \eeq
with solution
\beq \psi (t,\theta) \ = \ \exp[ - \, (m-1) \, s_2 \, \theta ] \ \eta (t) \ . \eeq

Equations \eqref{eq:deteq2a} and \eqref{eq:deteq2b} are thus satisfied, and we are now ready to tackle \eqref{eq:deteq1n}; this now reads -- factoring out an overall term $x^{-m} \exp[(m-1) s_2 z])$ -- as
\beq  2 \, x^2 \, \eta' (t) \ + \ \eta (t) \ \[ (m-1) \, (m \, s_1^2 \, x^{2 m} \ - \ s_2^2 \, x^2 ) \ + \ 2 \, m \, x \ f(x) \ - \ 2 \, x^2 \, \ f' (x) \]  \ = \ 0 \ . \eeq
The equation can be separated, and provides
$$ \eta (t) \ = \ q \ e^{K t} $$ (we set as usual $q=1$), while $f$ must satisfy
$$ - \ f' (x) \ + \ \frac{m}{x} \ f(x) \ + \ \frac{m \, (m-1) \, s_1^2}{2} \ x^{2(m-1)} \ - \ \frac{(m-1) \, s_2^2}{2} \ + \ K \ = \ 0 \ . $$
This is solved by $f(x)$ of the form \eqref{eq:fms}, where $C$ is an arbitrary constant, and we have written the constant $K$ appearing in $\eta$ and hence in $\vphi$ as
$$ K \ = \ (m-1) \, Q \ . $$
The function $\vphi$ reads now as \eqref{eq:phims}.
The proof is complete. \EOP

\bigskip
The usual remark about this symmetry being a random standard symmetry, and thus leading to an equation which is integrable but not in Ito form, applies here.

\section{Integrating symmetric scalar Ito equations with two Wiener processes}
\label{sec:inttwo}

For the Ito equations with symmetries, we can proceed to integration via use of the Kozlov transformation. We will now discuss this for the equations identified in Section \ref{sec:symmtwo}.

\subsection{Equations with additive and multiplicative noise}

Let us consider equations of the form \eqref{eq:Itoam} with drift of the form \eqref{eq:fam}; as seen above, these admit a standard symmetry identified by \eqref{eq:phiam}.

We will now apply the Kozlov procedure in this setting. This requires to pass to the variable
\beq y \ = \ \Phi (x,t;w,z) \ = \ \int \frac{1}{\vphi} \ d x \ ; \eeq
in this case we get
\beql{eq:yam} y \ = \ \exp \[  - \( \b \ - \ \frac{s_2^2}{2} \) \, t \ - \ s_2 \, z \] \ x \ := \ \Phi (x,t;w,z) \ . \eeq
Ito rule implies that
\beq d y \ = \ \( \frac{\pa \Phi}{\pa t} \) \, dt \ + \ \( \frac{\pa \Phi}{\pa x} \) \, dx \ + \ \( \frac{\pa \Phi}{\pa w} \) \, dw \ + \ \( \frac{\pa \Phi}{\pa z} \) \, dz \ + \ \frac12 \ \Delta (\Phi) \ dt \ . \eeq
Computing the derivatives and the Ito Laplacian, and substituting for $dx$ according to \eqref{eq:Itoam}, we obtain \beq d y \ = \ \exp \[  - \( \b \ - \ \frac{s_2^2}{2} \) \, t \ - \ s_2 \, z \] \ \( \a \, d t \ + \ s_1 \, d w \) \ . \eeq
This is of the form
\beql{eq:dyn} d y \ = \ F (t;w,z) \, dt \ + \ S_1 (t;w,z) \, dw \ + \ S_2 (t;w,z) \, d z \ , \eeq
actually with
\begin{eqnarray*}
F &=& \a \ \exp \[  - \( \b \ - \ \frac{s_2^2}{2} \) \, t \ - \ s_2 \, z \] \ , \\
S_1 &=& s_1 \ \exp \[  - \( \b \ - \ \frac{s_2^2}{2} \) \, t \ - \ s_2 \, z \] \ , \\
S_2 &=& 0 \ , \end{eqnarray*}
and is promptly integrated, see Section \ref{sec:symint}.


\subsection{Equations with Poisson and multiplicative noise}

In the case of Ito equations with Poisson and multiplicative noise, i.e. \eqref{eq:Itopm} with drift of the form \eqref{eq:fpm}, we have seen that the standard symmetry is identified by \eqref{eq:phipm}.

In this case we get
\beq y \ = \ \Phi (x,t;w,z) \ = \ 2 \ \sqrt{x} \ \exp \[ - \frac14 \ \( (2 \, c_3 \ - \ s_2^2) \, t \ + \ 2 \, s_2 \, z \) \] \ . \eeq
Proceeding as above, we get an equation of the form \eqref{eq:dyn}, now with
\begin{eqnarray*}
F &=& c_2 \ \exp \[ - \frac14 \ \( (2 \, c_3 \ - \ s_2^2) \, t \ + \ 2 \, s_2 \, z \) \]   \ , \\
S_1 &=& s_1 \ \exp \[ - \frac14 \ \( (2 \, c_3 \ - \ s_2^2) \, t \ + \ 2 \, s_2 \, z \) \] \ , \\
S_2 &=& 0 \ . \end{eqnarray*}
These are again directly integrated, as discussed in Section \ref{sec:symint}.

\subsection{Equations with multiplicative and general simple noise}

We can finally consider equations with multiplicative and simple noise (say with $m \not= 0,1/2,1$ to discard the trivial case and those already seen above), i.e. equations of the form  \eqref{eq:Itoms}. In this case symmetric equations are characterized by drift of the form \eqref{eq:fms} -- with $c$ and $Q$ arbitrary constants -- and the symmetry is identified by \eqref{eq:phims}.

In this case we get
\beq y \ = \ \Phi (x,t;w,z) \ = \ \frac{x^{1-m}}{1-m} \ \exp \[ (1 \, - \, m ) \ ( Q \, t \ - \ s_2 \, w )  \] \ . \eeq
Proceeding with the Kozlov substitution as above, we get an equation of the form \eqref{eq:dyn}, now with
\begin{eqnarray*}
F &=& C \ \exp \[ (1 \, - \, m ) \ ( Q \, t \ - \ s_2 \, w )  \]   \ , \\
S_1 &=& s_2 \ \exp \[ (1 \, - \, m ) \ ( Q \, t \ - \ s_2 \, w )  \] \ , \\
S_2 &=& 0 \ . \end{eqnarray*}

\section{Standard form of Ito equations, additive noise and change of variables}
\label{sec:oneconst}

As we have shown in Corollary 1, if one of the noises is additive and the other is simple, there is no symmetry, as we have shown; however, if we have a simple noise and a multiplicative one, we could get a symmetry, at least if the compatibility equation is satisfied.

On the other hand, it is well known (see e.g. \cite{Oksendal}) that we can transform any equation (even those with a multiplicative noise) into one with an additive noise\footnote{In fact, given the equation $d \xi = \phi (\xi,t) dt + \s (\xi,t) dw$, consider the variable $x = \int (1/\s (\xi,t)) d \xi$. This satisfies $d x = F(x,t) dt + dw$, where $F$ is a function which can be easily determined by Ito calculus.}. The problem is that in these cases no symmetry should exist.

This leads apparently to a contradiction in our results. In the present Section we will discuss this point; we will show that there is no contradiction, and actually our discussion will provide a different approach -- through the reduction of the Ito equation to a \emph{standard form}, in which one of the noises has a standard (usually additive) form -- to the analysis of Ito equations with two noises, confirming our results.

\subsection{Relation between equations with multiplicative and additive noise}

Consider the equation
\beql{eq:ns.1} dx \ = \ f(x) \, dt \ + \ s_1 \, x^k \, dw \ + \ s_2 \, dz \ , \ \ \ ( k \not= 1 ) \ ;
\eeq
As shown above, there is no symmetry. On the other hand, the equation
\beql{eq:ns.2}
dx \ = \ f(x) \, dt \ + \ s_1 \, \sqrt{x} \, dw \ + \ s_2 \, x \,  dz
\eeq
has a symmetry for some suitable form of the drift. The (apparent) problem is that we can change this equation \eqref{eq:ns.2} into an equation in which $z$ is an additive noise, i.e. has the form met in \eqref{eq:ns.1} (as we will see in a moment, this is only apparent due to what happens to the $w$ noise). In fact, consider the change of variable
\beq y \ = \ \Phi(x) \ = \ \int \frac{1}{x} \, dx \ = \ \log x \ .
\eeq
Then
$$  \frac{\pa \Phi}{\pa x} \ = \  \frac{1}{x} \ , \ \
\frac{\pa^2 \Phi}{\pa x^2} \ = \ - \, \frac{1}{x^2 } \ ,
$$
and we get
$$ dy \ = \ \left[ \frac{f(x)}{x} \ - \ \frac{1}{2\, x } \ \left( s_1^2 \ + \ s_2^2 \, x \right) \right] \, d t \ + \ \frac{s_1}{\sqrt{x}} \, dw \ + \ s_2 \, dz  \ . $$
The r.h.s. should be expressed in terms of $y$; this is obtained by writing $x=e^y$. That is, we get
\begin{eqnarray}
dy &=& \left[ \frac{f(e^y)}{e^y} \ - \ \frac{1}{2\, e^y } \ \left( s_1^2 \ + \ s_2^2 \, e^y \right) \right] \, d t \ + \ \frac{s_1}{\sqrt{e^y}} \, dw \ + \ s_2 \, dz \nonumber \\
& & := \ \wt{f} (y) \, dt \ + \ \wt{\s}_1 \, dw \ + \ \wt{\s}_2 \, dz  \ . \end{eqnarray}
Here we have defined
$$ \wt{\s}_1 \ = \ \frac{s_1}{\sqrt{e^y}} \ , \ \ \wt{\s}_2 \ = \ s_2 \ . $$
This explicit computation shows that passing to the $y$ variable the $z$ noise (which was multiplicative for the $x$ variable) is additive; but also that the $w$ noise is now \emph{not} a simple one.

Thus there is no contradiction with our previous results.
This simple discussion, however, calls for a look at the situation where we have an additive noise and a non-simple noise.

\subsection{Additive noise and non-simple noise}

Consider an Ito equation with an additive noise and a generic (in principle) noise:
\beq\label{onebis}
d x \ = \  f(x) \, dt \ + \ \sigma(x) \, dw \ + \ s_2 \, dz
\eeq
Note that if $\sigma (x)$ is linear we are in the case of a multiplicative noise; we will thus discard this possibility, and assume $\s (x) \not= s x$.

If we write (see also Appendix \ref{app:FODE})
\beq
b(x)\ = \ f(x) \ - \frac12 \, \sigma (x) \, \sigma' (x)
\eeq
the determining equations providing the function $\varphi$ for this equation are (recall we disregard W-symmetries in this section, as in previous ones):
\begin{eqnarray}
\varphi_t & \ = \  &  b'(x) \, \varphi \ - \ b(x)\, \varphi_x \ ; \label{three}\\
\varphi_w & \ = \  & \sigma'(x) \, \varphi \ - \ \sigma(x) \, \varphi_x \ , \label{four}\\
\varphi_z & \ = \  &\ - \ s_2 \, \varphi_x  \label{five} \ .
\end{eqnarray}

Assuming that equations \eqref{eq:deteq2a} and \eqref{eq:deteq2b} are satisfied, the functions
$$ J(x)\ = \ \frac{\sigma \, \sigma'' \ - \ \sigma'^2}{\sigma'} \ , \ \
K(x) \ = \ s_2 \ \frac{\sigma''}{\sigma'} $$
should be independent from $x$ to allow the existence of a symmetry.
From the equation for $K$, in the generic case (that is, if $\sigma$ is not linear), we have
$$ \sigma(x) \ = \ \alpha \, e^{\gamma x} \ + \ \beta \ , $$ with $\alpha$ and $\beta$ real constants, and $\ga = K/s_2$.

The equation for $J$ yields then
$$ J \ = \ \beta \, \gamma
\ . $$
Equations \eqref{four} and \eqref{five} are now respectively
\begin{eqnarray}
\varphi_w & =  & \alpha \, \gamma \, e^{\gamma x} \, \varphi\ - \ \( \alpha \, e^{\gamma x} \ + \ \beta \) \ \varphi_x \ , \label{fourB} \\
\varphi_z & =  & \ - \ s_2 \, \varphi_x  \ . \label{fiveB}
\end{eqnarray}
The second of these, equation \eqref{fiveB}, yields immediately
$$
\varphi(x,t;w,z) \ = \ h(t,w,\zeta) \ , \ \ \  \zeta \ = \ x \ - \ s_2 \, z \ ;
$$
now the first, equation \eqref{fourB}, becomes
$$
h_w   \ = \ - \ \alpha \( h_{\zeta} \ - \ \gamma  \, h\) \ e^{\gamma x} \ - \ \beta h_{\zeta} \ .
$$
We have to require
\beq
h_{\zeta} \ - \ \gamma \, h \ = \ 0 \ , \ \ \ \ h(t,w,\zeta) \ = \ g(t,w) \, e^{\gamma \zeta}
\eeq
and then,
$$
g_w   \ = \  - \beta \, \gamma \,  g \ , \ \ \ \  g(t,w) \ = \  \eta(t) \, e^{- \beta \gamma w} \ .
$$
In conclusion, the solution to our equations is
\beq
\varphi(x,t;w,z) \ = \ \eta(t) \ \exp \[\gamma \ \( x \ - \ s_2 \, z \ - \ \beta \, w \) \] \ .
\eeq

We substitute this expression into \eqref{three} and get
\beq
\eta'(t) \ = \ ( b_x(x) \ - \ \gamma \, b(x) ) \, \eta(t) \ .
\eeq
The coefficient of $\eta$ should be a constant, $-\gamma\,\delta$, and this fixes $b(x)$:
$$
b_x \ - \ \gamma \, b \ = \ - \ \gamma \, \delta \ , \ \ \ \ b(x) \ = \ \epsilon\,e^{\gamma x} \ + \ \delta \ . $$
Then, the drift $f(x)$ is
\beq
f(x)\ =\ b(x) \ + \ \frac12 \, \sigma \, \sigma_x \ = \ \frac12 \, \alpha^2 \, \gamma \, e^{2\gamma x} \ + \ \chi \, e^{\gamma x}  \ + \ \delta \ ,
\eeq
where $\chi$ is another constant.

Finally $\eta(t)$ satisfies (the constant factor can be taken equal to 1):
\beq
\eta' \ + \ \gamma \, \delta \, \eta \ = \ 0 \ , \ \ \ \ \eta(t) \ = \ e^{- \gamma \, \delta \, t } \ .
\eeq

We summarize our results as follows:

\medskip\noindent
{\bf Lemma 6.} {\it
The Ito equation \eqref{onebis}
with $\s (x) \not= s x$ admits a standard (random) symmetry if and only if  $f(x)$ and $\s (x)$ are of the form
\begin{eqnarray*}
f(x) &=& \frac{1}{2} \, \alpha^2 \, \gamma \, e^{2 \gamma  x} \ + \ \chi \, e^{\gamma  x} \ + \
 \delta  \ , \\
\sigma(x) &=& \alpha \, e^{\gamma x} \ + \ \beta \ , \end{eqnarray*}
where $\alpha,\beta,\gamma,\delta,\chi$ are arbitrary constants.

In this case the symmetry vector field $X  =  \varphi\,\pa_x$ is identified by
\beql{eq:symmlemma6}
\varphi(x,t;w,z)\ = \ \exp\[\gamma \, \( x \ - \ s_2 \, z \ - \beta \, w  \ - \ \delta \, t\) \] \ .
\eeq
}

\section{The case of $N$ noises}
\label{sec:NNoises}

We have stated, back in Sect.\ref{sec:CC}, that the case of $N$ noises could be analyzed along the same lines as the case of two noises. We will now briefly discuss the case of $N$ noises, still keeping to the autonomous case and to standard symmetries; that is, we consider equations of the form
\beql{eq:Ito:Nnoises} dx \ = \ f(x) \, dt \ + \ \sum_{k=1}^N \ \s_{(k)} (x) \, d w^k \ . \eeq
We assume, to discard trivial cases, that all the $\s_{(k)} (x)$ are functionally independent.
Note that the index of $\s$ is set within brackets to distinguish it from a differentiation symbol.

\medskip\noindent
{\bf Lemma 7.} {\it The equation \eqref{eq:Ito:Nnoises} can admits a standard symmetry only if all the functions
$$ J_{(mn)} \ := \ \frac{\s_{(m)} \, \pa_x [\s_{(m)} \ (\s_{(n)})_x \ - \ (\s_{(m)})_x \ \s_{(n)}] \ - \ (\pa_x \s_{(m)} ) \, [\s_{(m)} \ (\s_{(n)})_x \ - \ (\s_{(m)})_x \ \s_{(n)}] }{\s_{(m)} \ (\s_{(n)})_x \ - \ (\s_{(m)})_x \ \s_{(n)}} $$
($m,n = 1,...,N$) are constant.}

\medskip\noindent
{\bf Proof.} For the equation \eqref{eq:Ito:Nnoises}, the determining equations for standard symmetries will be
\begin{eqnarray}
\vphi_t &+& f \, \vphi_x \ - \ f_x \, \vphi \ = \ - (1/2) \ \De (\vphi ) \ , \label{eq:NN1} \\
\vphi_k &+& \s_{(k)} \, \vphi_x \ - \ (\s_{(k)})_x \, \vphi_k \ = \  0 \ . \label{eq:NNk} \end{eqnarray}
It should be noted that now
$$ \De (\phi) \ := \ \sum_{k=1}^N \phi_{kk} \ + \ 2 \, \sum_{k=1}^N \s_{(k)} \, \phi_{xk} \ + \ \sum_{k=1}^N \s_{(k)}^2 \, \phi_{xx} \ . $$
The equation \eqref{eq:NN1} can also be recast in first order form as
\beq \vphi_t \ + \ b \, \vphi_x \ - \ b_x \, \vphi \ = \ 0 \ , \eeq
provided we define
\beq b (x) \ := \  f (x) \ - \ \frac12 \ \sum_{k=1}^N \s_{(k)} (x) \, \s_{(k)}' (x) \ . \eeq

Proceeding as in Sect.\ref{sec:symmtwo}, we consider the equations \eqref{eq:NNk} for any two values of the index $k$, say $k=m$ and $k = n$. By cross-differentiating and considering the difference of the two equations thus obtained, we get
\beq \( \s_{(m)} \ (\s_{(n)})_x \ - \ (\s_{(m)})_x \ \s_{(n)} \) \ \vphi_x \ = \ \( \s_{(m)} \ (\s_{(n)})_{xx} \ - \ (\s_{(m)})_{xx} \ \s_{(n)} \) \ \vphi \ . \eeq
Writing for short
\beql{eq:LAml} \La_{mn} (x) \ := \ \s_{(m)} \ (\s_{(n)})_x \ - \ (\s_{(m)})_x \ \s_{(n)} \ , \eeq
this equation yields (here and below, no sum on $m,n$)
\beql{eq:phiccml} \vphi (x,t;\wb) \ = \ \La_{(mn)} (x) \ h_{(mn)}(t;\wb)  \ . \eeq
Note that $\La_{(mn)} (x) \not= 0$ unless $\s_{(m)}$ and $\s_{(n)}$ are functionally dependent, which we have excluded. Inserting this expression in \eqref{eq:NNk}, we get
\beq \La_{mn} \ \pa_k h_{mn} \ + \ \( \s_{(m)} \, (\pa_x \La_{(mn)} ) \ - \ (\pa_x \s_{(m)} ) \, \La_{(mn)} \) \ h_{(mn)} \ = \ 0 \ , \eeq that is
\beql{eq:hmn} \frac{\pa_k h_{(mn)}}{h_{(mn)}} \ = \ \frac{\( (\pa_x \s_{(m)} ) \, \La_{(mn)} \ - \ \s_{(m)} \, (\pa_x \La_{(mn)} )  \)}{\La_{(mn)}} \ . \eeq
The r.h.s. of this equation is just the function $J_{(mn)}$ defined in the statement.

Now we observe that the $h_{(mn)}$ do not depend on the $x$ variable; thus the l.h.s. of \eqref{eq:hmn} has vanishing $x$ derivative, and the same holds for its r.h.s.; in other words, we have proved that the condition for the equations \eqref{eq:NNk} to admit a common solution is that
\beql{eq:Jcc} \pa_x \ J_{(mn)} \ = \ 0 \ \ \ \forall m,n = 1,...,N \ . \eeq
Note that this gives no information about common solutions to the equations \eqref{eq:NNk} for $k=1,...,N$ and the equation \eqref{eq:NN1}; thus we have a necessary but not sufficient conditions for the existence of a standard symmetry. This completes the proof. \EOP

\medskip\noindent
{\bf Corollary 2.} {\it For $N \ge 3$, the equation \eqref{eq:Ito:Nnoises} can not admit any standard symmetry if all the noise terms are simple ones, $\s_k (x) = s_k x^{j_k}$.}

\medskip\noindent
{\bf Proof.} In the case of simple noises, the compatibility condition \eqref{eq:Jcc} can be satisfied only if either $j_m = 1$ or $j_n = 1$; this follows from Corollary 1 in Sect.\ref{sec:CC}. As we assumed the noise coefficients $\s_k (x)$ are all functionally independent, at most one of them (say the one with $k=1$) can give a multiplicative noise, and hence all the $(N-1)(N-2)$ functions $J_{(mn)}$ with both $m \not= 1$ and $n \not= 1$ are non-constant. \EOP
\bigskip

We note that our argument does not apply to non-simple noises. However, it turns out the same result holds in full generality.

\medskip\noindent
{\bf Lemma 8.} {\it For $N \ge 3$ and independent noises, the equation \eqref{eq:Ito:Nnoises} can not admit any standard symmetry.}

\medskip\noindent
{\bf Proof.} It is clear that it suffices to prove the statement for $N=3$. We note preliminarily that -- by the standard transformation discussed above, see Section \ref{sec:oneconst} -- we can always reduce to the case where one of the noises, say the third one, is an additive one (actually with unit coefficient).

In this case, the statement follows directly from the determining equations. These determine both the drift and the noise coefficients for the first two noises. The equations can be solved explicitly, and it turns out that if a standard symmetry exists, then the noises are necessarily \emph{linearly dependent}. We refer to Appendix \ref{app:3N} for full detail of the required computations. \EOP

\section{Discussion and conclusions}
\label{sec:discussion}

The classification of scalar autonomous Ito equation with a single noise source which admit a standard symmetry and which are hence integrable via the Kozlov substitution has been recently completed and is available in the literature \cite{Koz1,Koz2,Koz3,GS20,GR22}.

We have noted that in many physical situation a system may be subject to \emph{multiple} noise sources: e.g. environmental and demographic noise for population dynamics, production and shot noise in electronics, etc.
With this motivation, and also for its purely mathematical interest, we have considered scalar autonomous Ito equations with \emph{two} different -- and independent -- noise sources. It appears that the methods developed in this case could be readily extended to an arbitrary number of independent noise sources, but we limited our investigation to the case of two sources, and considered Ito equations of the form  \eqref{eq:Itotwo}.

First of all, we formulated a compatibility condition between the coefficients of the two noises, see Section \ref{sec:CC}; a symmetry can be present only if this is satisfied, see \emph{Lemma 1}. It is a simple consequence of this that, in the case of simple noises, a symmetry can be present only if one of the noises is a multiplicative one, see \emph{Corollary 1}.

We have then considered in detail the combinations of noises which appear more frequently in applications, physical and otherwise, and for each of these we have identified the equations admitting symmetries, and determined their symmetries. Our results are summarized in \emph{Lemma 2} for the combination of additive and multiplicative noise, in \emph{Lemma 3} (which confirms in an independent way what was expected from the general result in \emph{Lemma 1}) for the combination of additive and of Poisson noise, in \emph{Lemma 4} for the combination of additive and multiplicative noises. We also considered the combination of a multiplicative and of a general simple noise -- that is a noise term $s x^m$ -- and the result for this case is given in \emph{Lemma 5}.

As mentioned above, knowledge of a standard symmetry allows to integrate a scalar Ito equation; this is done through an explicit change of variable, known as the Kozlov substitution, see Section \ref{sec:inttwo}. In the case of a \emph{deterministic} standard symmetry this maps the symmetric equation into a new (integrable) Ito equation, while in the case of a \emph{random} standard symmetry the equation is mapped into an integrable equation which is \emph{not} in Ito form.

The standard symmetries identified by our study, see \emph{Lemmas 2, 4 and 5}, are actually random standard symmetries, so we expect the symmetric equations will be mapped into non-Ito integrable equations, and this is indeed the case.

We have then considered in Section \ref{sec:inttwo} this explicit map and determined the integrable form of the different symmetric equations.

In Section \ref{sec:oneconst} we discussed an alternative approach, based on the reduction of the Ito equation to a standard form -- in which one of the two noises is an additive one -- to check our results reobtaining them in a different way; this also allowed to emphasize and better understand the role of the restriction to simple noises for some of our results, stressing once more they hold for simple noises.

Finally, in Section \ref{sec:NNoises} we have discussed the case of $N$ noises, and show that for an autonomous equation with $N$ independent noises, be these simple (\emph{Lemma 7}) or of general form (\emph{Lemma 8}), there is no symmetry for $N > 2$, see \emph{Lemma 8}. In other words, for scalar equations with simple noises the present work provides a complete classification.

We trust our work and results will be of help to anybody dealing with stochastic equations in which several noise sources are present. Of course the case of integrable equations is a very special one, and we expect that it occurs in a very limited number -- if any -- of applications. But already knowing that the problem at hand is not integrable can be a help, saving efforts; and moreover, as in the case of deterministic equations, equations which are near enough to an integrable one can be treated via a perturbation theory approach.

The work is completed by three Appendices. In Appendix \ref{app:FODE} we recall, following our previous work \cite{GR22}, how the determining equations for standard symmetries of an Ito equation (which are in principle a set of second order PDEs) can be cast in first order form. In Appendix \ref{app:NOW} we show that the restriction to standard symmetry, disregarding so called proper W-symmetries \cite{GSW}, was indeed harmless, in that these cannot be used for integration. In Appendix \ref{app:3N} we give full detail of the computations showing that an autonomous equation with three independent noises can not admit a standard symmetry.

\addcontentsline{toc}{subsection}{Acknowledgements}

\subsection*{Acknowledgements}

We are grateful to the anonymous Referee for a careful checking of our paper. This work is a fallout of a semester-long stay of MAR in Milano under the Program ``Salvador de Madariaga''. The support of Universidad Complutense de Madrid under grant G/6400100/3000, is also acknowledged. The paper was written while GG was in residence at, and enjoying the relaxed atmosphere of, SMRI. The work of GG is also supported by GNFM-INdAM.

\newpage

\begin{appendix}

\section{Determining equations in first order form}
\label{app:FODE}

The determining equations for symmetries of an Ito equation with a single noise source can be cast in first order form -- obtaining the determining equations for the symmetries of the associated Stratonovich equation -- as discussed in detail in our recent paper \cite{GR22}. Here we discuss how the same approach can be applied to the present case of multiple noise sources; we will actually limit, for the sake of simplicity, to the case of two noise sources; but generalization to $N$ noise sources is a simple matter.

Let us thus consider standard symmetries for an Ito equation with two different noises, $\sigma (x)$ and $\rho(x)$. These have to satisfy the determining equations
\begin{eqnarray}
& & \vphi_t \ + \ f \, \vphi_x \ - \ \vphi \, f_x \ + \ \frac12 \, \Delta (\vphi ) \ = \ 0 \ , \label{eq:app:deteq1gen} \\
& & \vphi_{w} \ + \ \s  \, \vphi_x \ -  \ \s_x \, \vphi  \ = \ 0  \ , \label{eq:app:deteq2gen}\\
& & \vphi_{z} \ + \ \rho \, \vphi_x \ - \ \rho_x \, \vphi \ = \ 0  ; \label{eq:app:deteq3gen}
\end{eqnarray}
where $\Delta$ is the  Ito Laplacian, which in this case reads
\beql{eq:app:Delta} \Delta (\phi) \ := \ \phi_{ww} \ + \ \phi_{zz} \ + \ 2 \, ( \s \, \phi_{xw} +\rho \, \phi_{xz})\ + \ ( \s^2 \ + \ \rho^2)\ \phi_{xx} \ . \eeq

We can proceed as in the single noise case, converting  equation \eqref{eq:app:deteq1gen} into a first order equation using equations \eqref{eq:app:deteq2gen} and \eqref{eq:app:deteq3gen}. In fact, these provide
\begin{eqnarray*}
\vphi_w & = &  \s_x \, \vphi \ -\ \s  \, \vphi_x \ , \\
\vphi_z & = &   \rho_x \, \vphi \ - \ \rho  \, \vphi_x \ .
\end{eqnarray*}
Differentiating them, we obtain
\begin{eqnarray*}
\vphi_{xw} & = &  \s_{xx} \, \vphi \ -\ \s  \, \vphi_{xx}  \ , \\
\vphi_{ww} & = &  (\s_{x}^2 \ -\ \s \, \s_{xx})\, \vphi \ - \ \s \,\s_x \, \vphi_{x} \ + \ \s^2\, \vphi_{xx} \ , \\
\vphi_{xz} & = &  \rho_{xx} \, \vphi \ -\ \rho  \, \vphi_{xx}  \ , \\
\vphi_{zz} & = &  (\rho_{x}^2 \ -\ \rho \, \rho_{xx})\, \vphi \ - \ \rho \,\rho_x \, \vphi_{x} \ + \ \rho^2\, \vphi_{xx} \ .
\end{eqnarray*}
Substituting in \eqref{eq:app:Delta} we get
\beq
\Delta\,\vphi \ = \ ( \s_x^2 \ + \ \rho_x^2 \ + \ \s \, \s_{xx} \ + \  \rho \, \rho_{xx}  ) \, \vphi \ - \ ( \s \s_x\ +\ \rho\, \rho_x ) \, \vphi_x \ .
\eeq

Equation \eqref{eq:app:deteq1gen} is hence written as:
\beq
\vphi_t \ +\ \left(\ f-\frac14(\ \s^2\ + \ \rho^2\ )_x\right)\ \vphi_x \ - \ \left(\ f-\frac14(\ \s^2\ + \ \rho^2\ )_x  \right)_x \ \vphi \ = \ 0 \ .
\eeq
With the definition
$$ b \ := \ f \ - \ \frac14 \( \s^2 \ + \ \rho^2 \)_x \ ,
$$
see \eqref{eq:btwodim}, it reads
\beq
\vphi_t \ + \ b \, \vphi_x \ - \ b_x   \, \vphi \ = \ 0 \ .
\eeq
This is the same as \eqref{eq:deteq1n}.

We note that the same approach also allows to express in first order form the determining equations for W-symmetries, see \eqref{eq:deteq1nso}, \eqref{eq:deteq2ar}, \eqref{eq:deteq2br}. In this case solving \eqref{eq:deteq2ar} and \eqref{eq:deteq2br} for $\vphi_w$ and $\vphi_z$ respectively yields
\begin{eqnarray*}
\vphi_w &=& \vphi \, \s_x \ - \ \s \, \vphi_x \ + \ r_{11} \, \s \ + \ r_{12} \vr \ ,  \\
\vphi_z &=& \vphi \, \vr_x \ - \ \vr \, \vphi_x \ + \ r_{21} \, \s \ + \ r_{22} \, \vr \ ; \end{eqnarray*}
differentiating these we obtain expressions for $\vphi_{xw},\vphi_{ww},\vphi_{xz},\vphi_{zz}$; these can be substituted in the explicit expression for $\De (\vphi)$, see \eqref{eq:Deltan}, within \eqref{eq:deteq1nso}. In this way the latter reads
\beql{WsymmFO} \vphi_t \ + \ b (x,t) \ \vphi_x \ - \ b_x (x,t) \ \vphi \ + \ \ga (x,t) \ = \ 0 \ , \eeq where $b(x,t)$ is the Stratonovich drift \eqref{eq:btwodim} for our equation. Equation \eqref{WsymmFO} is indeed a first order equation for $\vphi$; together with \eqref{eq:deteq2ar} and \eqref{eq:deteq2br} this forms a first order system.

As for $\ga (x,t)$, its explicit expression is
\beq \ga \ = \  \( r_{11} \, \s \, \s_x \ + \ r_{22} \, \vr \, \vr_x \) \ + \ \frac12 \
\( r_{12} \ + \ r_{21} \)  \ \( \s \, \vr_x \ + \ \s_x \, \vr \) \ . \eeq

We have thus recovered eq.\eqref{eq:deteq1nfor}.


\section{On W-symmetries and integration of Ito equations}
\label{app:NOW}

We chose to consider only \emph{standard symmetries}, and disregard so called \emph{W-symmetries}. The main reason for this choice is that W-symmetries do not lead to integration of the Ito equation (see the Introduction in this regard).

In this Appendix we will briefly justify this statement and thus our choice; we will just discuss the simpler case of a scalar Ito equation depending on a single noise,
\beql{eq:na.Ito} d x \ = \ f (x,t) \, dt \ + \ \s (x,t) \, d w \ . \eeq
A W-symmetry has generator $X = \vphi(x,t,w) \pa_x + r w \pa_w$; but we will only consider the even simpler case of \emph{split W-symmetries} \cite{GSW}, i.e.
\beq X \ = \ \vphi (x,t) \, \pa_x \ + \ r \, w \, \pa_w \ . \eeq
In order to integrate the Ito equation \eqref{eq:na.Ito}, we should find a change of variables
$$ (x,t;w) \ \to \ (y, \tau ; z ) $$
such that in the new variables the symmetry vector field read as
$$ X \ = \ \pa_y \ . $$
This requires to determine functions $y = \psi (x,t;w)$, $\tau = \theta (x,t;w)$ and $z = \zeta (x,t;w)$ such that
\beq X (\psi ) \ = \ 1 \ , \ \ \ X (\theta ) \ = \ 0 \ , \ \ \ X (\zeta ) \ = \ 0 \ . \eeq
It is immediately apparent that the simplest solution for $\theta$, which we will take, is
$$ \tau \ = \ t \ . $$
Moreover, again in the search for simple solutions, we can choose $\psi$ and $\zeta$ to be independent of $t$; i.e. assume $\psi = \psi (x,w)$, $\zeta = \zeta (x,w)$.

Now the problem amounts to solving two linear PDEs for $\psi$ and $\zeta$,
\begin{eqnarray*}
\vphi \, \frac{\pa \psi}{\pa x} \ + \ r \, w \, \frac{\pa \psi}{\pa w} &=& 1 \ , \\
\vphi \, \frac{\pa \zeta}{\pa x} \ + \ r \, w \, \frac{\pa \zeta}{\pa w} &=& 0 \ . \end{eqnarray*}
These can be tackled by the method of characteristics. Denoting by $\chi = \chi (x,w)$ the characteristic variable for the vector field $X$ (note that here we take advantage of our restriction to split W-symmetries, i.e. of the assumption that $\vphi$ does not depend on $w$),
\beq \chi \ = \ w \ \exp \[ - \ \int \frac{r}{\vphi (x,t)} \ dx \] \ , \eeq
the solutions to our system are given by
\begin{eqnarray}
\psi (x,t;w) &=& \int \frac{1}{\vphi (x,t)} \ dx \ + \ \a (\chi,t) \ , \\
\zeta (x,t;w) &=& \b (\chi,t) \ . \end{eqnarray}
The requirement that these function define a proper change of variables rules out the possibility to choose $\b \equiv 0$, or more generally $\b = const$ (note that we can instead choose $\a \equiv 0$). Thus the simplest choice is
$$ \b (\chi,t) \ = \ \chi \ . $$

In any case, it follows from the previous general formulas that the new random variable $z = \zeta (x,t;w)$ is \emph{not} a Wiener process, and actually its statistical properties depend on the process $x(t)$ (that is, on the solution to our SDE) itself.

In other words, albeit we will in this way manage to write our equation in the form
\beq d y \ = \ F (t;z) \, dt \ + \ S(t;z) \, dz \ , \eeq
not only this equation will not be in Ito form (since the drift and the noise coefficient will depend on the driving process $z$), but moreover the process $z(t)$ will not be a Wiener one.

Thus, in general, W-symmetries -- even in the simplest case of split W-symmetries of a scalar equation -- can not be used to integrate a Ito equation.

We have at several points chosen the simplest solution; one could -- and maybe should -- wonder if a less simple one would have led to different conclusions, and this point is worth a brief discussion. First of all, we note that the general solution for $X(\theta) = 0$ would have been $\theta = h (t,\chi)$. But a dependence of $\tau$ on $(x,w)$ would mean the time is changed to a random variable itself; this should not be allowed, as discussed in \cite{GRQ1,GGPR,GSW}. One could allow $\tau = h(t)$ (with $h$ a monotone function, see again \cite{GRQ1,GGPR,GSW}), which leads to a reparametrization of time. This would produce some more involved formulas, but no substantial change of our point. Similarly, once we set $\tau = h(t)$ we can not take $\z$ independent of $\chi$, or the Jacobian of the change of variables would be singular. Once there is a dependence of $\z$ on $\chi$, the new driving process is necessarily not a Wiener one. So the essence of our discussion holds also in the more general case.

\section{Non existence of symmetric equations with three independent noises}
\label{app:3N}

In this Appendix we give full detail of the computations leading to the result stated in Lemma 8, i.e. the impossibility of equations with three (and hence also of more) \emph{independent} noises admitting a standard symmetry.

Consider an autonomous equation with three noises
\beql{eq:threenoises} d \wt{x} \ = \ \wt{f} (\wt{x} ) \ dt \ + \ \sum_{i=1}^3 \wt{\s}_i (\wt{x} ) \ d w_i \ . \eeq
By the standard change of variables
$$ x \ = \ \int \frac{1}{\s_3 (\wt{x} )} \, d \wt{x} \ , $$ see Section \ref{sec:oneconst}, we can always get one of the three noise coefficients, in this case the one for $w_3$, to be a constant -- so that $w_3$ is an additive noise. That is, we get $\s_3 (x) = c$; we can in fact always take $c = 1$ (computations would be absolutely the same, with some small notational complication,  for generic $c$).
Thus, we can always reduce to study
\beq d x \ = \ f (x)  \, dt \ + \ \s_1 (x) \, d w_1 \ + \ \s_2 (x) \, d w_2 \ + \ d w _3 \ . \eeq
Moreover, for ease of notation,  we write
$$ \s_1 (x) \ = \ \s (x) \ , \ \ \s_2 (x) \ = \ \rho (x) \ ; \ \ \ \vphi_k := \pa \vphi / \pa w_k \ . $$ The determining equations for standard symmetries $X = \vphi (x,t;w_1,w_2,w_3) \pa_x $ are
\begin{eqnarray}
\vphi_t \ + \ b \, \vphi_x \ - \ \vphi \, b_x &=& 0 \ ; \label{deteq0} \\
\vphi_k \ + \ \s_k \, \vphi_x \ - \ \vphi \, (\s_k)_x &=& 0 \ \ \ (k=1,2,3) \ . \label{deteqK} \end{eqnarray}
We will now study and solve these.

The equation \eqref{deteqK} for $k=3$ yields
\beql{eq:k3} \vphi (x,t;w_1,w_2,w_3) \ = \ \chi (t;w_1,w_2,z) \ , \ \ \ z \ := \ w_3 \ - \ x \ . \eeq
From now on we will thus set $ w_3 \equiv x + z$.

Looking now at the equation \eqref{deteqK} for $k=2$, we have
\beql{eq:k2} \frac{\pa \chi}{\pa w_2} \ - \ \rho \ \frac{\pa \chi}{\pa z} \ - \ \rho' \ \chi \ = \ 0 \ . \eeq

Differentiating this in $x$, we get
$$ \frac{\rho''}{\rho'} \ = \ - \ \frac{\chi_z}{\chi} \ ; $$
as the functions on the two sides depend on different sets of variables, this requires
$$
\frac{\rho''}{\rho'} \ = \  \alpha \ = \ - \ \frac{\chi_z}{\chi}  $$
with $\alpha$ a constant. These equations yield
\beq
\rho (x) \ = \ k_2 \ e^{\alpha x} \ + \ c_2 \ , \ \ \ \
\chi (t,w_1,w_2,z) \ = \ e^{- \alpha z} \ \^\chi (t,w_1,w_2) \ . \eeq
Recall that this follows from a differential consequence of \eqref{deteqK} for $k=2$; when we look at the equation itself, we now get
$$  e^{- \a z} \ \( \frac{\pa \^\chi}{\pa w_2} \ + \ \a \, c_2 \ \^\chi \) \ = \ 0 \ ; $$
this yields promptly
\beql{chihat} \^\chi (t,w_1,w_2) \ = \ e^{- \a \, c_2 \, w_2} \ \psi (t,w_1) \ . \eeq

We now pass to consider equation \eqref{deteqK} with $k=1$, which reads
$$ e^{- \a (c_2 w_2 + z)} \ \[ \( \a \, \s \ - \ \s' \) \ \psi \ + \ \frac{\pa \psi}{\pa w_1} \] \ = \ 0 \ . $$
Differentiating this in $x$, we get
$$  e^{- \a (c_2 w_2 + z)} \ \psi \ \( \a \, \s' \ - \ \s'' \) \ = \ 0 \ . $$
Requiring $\psi \not= 0$ (or we would have no symmetry), this requires to have
$  \a \s'  =  \s''$, which of course implies
\beq \s (x) \ = \ k_1 \ e^{\a x} \ + \ c_1 \ . \eeq
Again, this follows from a differential consequence of \eqref{deteqK} for $k=1$, and we should look at the equation itself, which now reads
$$  e^{- \a ( c_2 w_2 + z) }  \ \( \frac{\pa \psi}{\pa w_1} \ + \ \a \, c_1 \, \psi \) \ = \ 0 \ . $$
Solving this, we get
\beq \psi (t,w_1) \ = \ e^{- \a \, c_1 \, w_1 } \ \eta (t) \ . \eeq

Finally, we should now look at the first determining equation \eqref{deteq0}. This now reads
$$ \frac12 \ e^{- \a (c_1 w_1 + c_2 w_2 + z)} \ \[ 2 \, \eta' (t)  \ + \ \eta (t) \ \( \a^2 \ e^{2 \a x} \, (k_1^2 + k_2^2 ) \ + \ 2 \, \a \, f (x) \ - \ 2 \, f' (x)  \) \] \ = \ 0 \ . $$
Differentiating this in $x$, and omitting the overall (never vanishing) exponential term, we get
$$ \eta \ \[ \a^3 \ e^{2 \a x} \ (k_1^2 + k_2^2) \ + \ \a \, f' \ - \ f'' \] \ = \ 0 \ . $$ As we do not want to have $\eta = 0$ (this would imply $\vphi = 0$), we have to solve
$$  \a^3 \ e^{2 \a x} \ (k_1^2 + k_2^2) \ + \ \a \, f' \ - \ f'' \ = \ 0 \ , $$
and this yields (with $\gamma$ and $\delta$ arbitrary constants)
\beql{eq:f3N} f(x) \ = \ \frac12 \ \a \, (k_1^2 + k_2^2 ) \ e^{2 \a x} \ + \ e^{\a x} \, \gamma \ + \ \delta \ . \eeq

Going back to equation \eqref{deteq0} itself, this now reads
$$  \exp \[ - \, \a \, \( c_1 w_1 \, + \, c_2 w_2 \, + \, z \) \] \ \( \eta' \ + \ \a \, \delta \, \eta \) \ = \ 0 \ . $$
The solution is provided (omitting as usual the overall multiplicative constant) by
\beq \eta (t) \ = \ \exp [ - \, \a \, \delta \, t ] \ . \eeq

Summarizing, we found that an equation of the form \eqref{eq:threenoises} admits a standard symmetry if and only if the drift coefficient $f(x)$ is of the form \eqref{eq:f3N}, \emph{and moreover the noise coefficients are of the forms we have determined}; in this case the symmetry $X = \vphi \pa_x$ is identified by
\beql{eq:phi3N} \vphi (x,t,w_1,w_2,w_3) \ = \ \exp \[ - \, \a \ \( \de t + c_1 w_1 + c_2 w_2 + w_3 - x \) \] \ . \eeq

However, we should look in more detail at the noise coefficients we have been determining in the course of our computation. These are
\begin{eqnarray*}
\s_1 (x) &=& \s (x) \ = \ c_1 \ + \ k_1 \ e^{\a x} \ , \\
\s_2 (x) &=& \rho (x) \ = \ c_2 \ + \ k_2 \ e^{\a x} \ , \\
\s_3 (x) &=& 1 \ . \end{eqnarray*}
It is immediate to check that these are linearly dependent.

More precisely, we have $q_1 \s_1 + q_2 \s_2 + q_3 \s_3 = 0 $
with $q_3 \not= 0$ an arbitrary number and
$$ q_1 \ = \ \frac{k_2}{c_2 \, k_1 \ - \ c_1 \, k_2} \ q_3 \ , \ \ \
q_2 \ = \ \frac{k_1}{c_1 \, k_2 \ - \ c_2 \, k_1} \ q_3 \ . $$

We conclude that it is not possible to have an (autonomous, scalar) Ito equation with three \emph{independent} noises which admits a standard symmetry.

\end{appendix}

\label{lastpage}
\end{document}